\documentclass[a4paper]{jpconf}
\usepackage{graphicx}

\newcommand{\PbPb}{\mbox{Pb--Pb}}
\newcommand{\pPb}{\mbox{p--Pb}}

\newcommand{\sqrtsNN}{\sqrt{s_{\rm \scriptscriptstyle NN}}}

\newcommand{\RpPb}{R_{\rm pPb}}
\newcommand{\DtoKpi}{{\rm D^0\to K^-\pi^+}}
\newcommand{\DtoKpipi}{{\rm D^+\to K^-\pi^+\pi^+}}
\newcommand{\DstartoDpi}{{\rm D^{*+}\to D^0\pi^+}}

\newcommand{\Dstophipi}{{\rm D_s^{+}\to \phi\pi^+}}

\newcommand{\Dzero}{{\rm D^0}}
\newcommand{\Dstar}{{\rm D^{*+}}}
\newcommand{\Dplus}{{\rm D^+}}
\newcommand{\Ds}{{\rm D_s^+}}

\begin{document}
\title{Measurement of D-meson production in p-Pb collisions with the ALICE detector}

\author{Grazia Luparello (for the ALICE Collaboration)}

\address{Nikhef, Science Park 105, 1098 XG Amsterdam, The Netherlands}

\ead{E-mail: Grazia.Luparello@cern.ch}

\begin{abstract}
The ALICE Collaboration has measured the production of prompt D mesons in $\pPb$ collisions at $\sqrtsNN = 5.02$ TeV in the rapidity range $-0.04 < y_{\rm cms} < 0.96$ via the exclusive reconstruction of their hadronic decays: $\DtoKpi$, $\DtoKpipi$, $\DstartoDpi$ and $\Dstophipi$. The $p_{\rm {T}}$-differential production cross sections and the $p_{\rm {T}}$-dependent nuclear modification factors with respect to a proton-proton reference, $\RpPb$, are presented. 
\end{abstract}

\section{Introduction}
Proton-nucleus interactions at the LHC allow for the study of fundamental properties of Quantum Chromodynamics (QCD) at low parton fractional momentum $x$ and high gluon densities \cite{bib:Salgado_pPb} and they provide a reference for the studies of deconfined matter created in nucleus-nucleus collisions. 
Measurements of charm particle production in central Pb-Pb collisions at $\sqrtsNN~=~2.76$~TeV showed a suppression of the D-meson yield with respect to the binary-scaled pp yield in a broad $p_{\rm T}$ range \cite{bib:PbPbDmeson276}. This effect is consistent with the energy loss of charm quarks while they traverse the hot and dense medium formed in such collisions. To come to a quantitative understanding of the energy loss results, it is important to disentangle hot medium effects from initial-state effects due to cold nuclear matter, such as a modification of the parton distribution functions in the nucleus \cite{bib:EPS09} and gluon saturation effects \cite{bib:CGC}. Initial-state effects can be isolated by studying the D meson production in p-Pb collisions, where a hot medium is not expected to form. A sensitive observable is the nuclear modification factor, $\RpPb$.  In case of minimum bias $\pPb$ interactions $\RpPb$ is defined as
$R_{\rm pPb}= \frac{\rm 1}{\rm A} \frac{{\rm d} \sigma_{\rm pPb} / {\rm d}p_{{\rm T}}}{{\rm d} \sigma_{\rm pp}/ {\rm d}p_{{\rm T} }}$, where A is the number of nucleons in the lead nucleus, $\sigma_{\rm pp}$ and $\sigma_{\rm pPb}$ are the production cross sections of D mesons in pp and $\pPb$ collisions respectively. In these proceedings, we present the results obtained from a data sample of $\sim$100M $\pPb$ collisions at $\sqrtsNN = 5.02$ TeV  collected with a Minimum Bias trigger defined in ALICE by the coincidence of signals in the two VZERO detectors, consisting of two arrays of scintillator counters covering the pseudorapidity regions $2.8<\eta<5.1$ and $-3.7<\eta<-1.7$.

\section{Measurement of D-meson production in $\pPb$ collisions}
The analysis strategy is based on the invariant mass analysis of fully reconstructed decay topologies. $\Dzero$, $\Dplus$, $\Dstar$ and $\Ds$ mesons and their antiparticles are reconstructed via their hadronic decay channels $\DtoKpi$ (with branching ratio, BR, of $3.88 \pm 0.05$ \%), $\DtoKpipi$ (BR of $9.13 \pm 0.19$ \%), $\DstartoDpi$ (BR of $67.7 \pm 0.5$ \%) and $\Dstophipi$ with the subsequent decay $\phi \rightarrow {\rm K^{+} K^{-}}$ (BR of $2.28 \pm 0.12$ \%) \cite{bib:PDG}. The extraction of the D-meson signals out of the large combinatorial background from uncorrelated tracks is based on the reconstruction and selection of secondary vertex topologies with a separation of a few hundred micrometers from the primary vertex. The Time Projection Chamber (TPC) and the Inner Tracking System (ITS) detectors, in particular its two innermost layers equipped with silicon pixel detectors, provide a resolution of the track impact parameter of about 70~$\mu$m for tracks with $p_{\rm T}=1$~GeV/$c$. To further suppress the combinatorial background, particle identification (PID) on the D-meson decay products is employed. PID is performed using the information on specific energy deposit in the TPC and on the time of flight measured with the Time Of Flight (TOF) detector. 
The signal yield is extracted by fitting the invariant mass distribution using a Gaussian function for the signal peak. The background is fitted using an exponential shape in the case of $\Dzero$, $\Dplus$ and $\Ds$. For the $\Dstar$, the distribution of $\Delta {\rm M} = {\rm M} ({\rm K^-} \pi^+ \pi^+) - {\rm M} ({\rm K^-} \pi^+)$ is fitted with a threshold function convoluted with an exponential. The raw yields extracted by the fit are divided by 2 to give the average (particle and antiparticle) cross section.
The correction for acceptance and efficiency is performed using Monte Carlo simulations based on PYTHIA with the Perugia-0 tuning to generate the D-meson signals on top of an underlying event
generated with HIJING. The efficiencies are weighted to take into account the dependence of the D-meson production on the number of charged particles produced in the collisions.

In order to extract the cross section of prompt D mesons, the contribution of D mesons from B decays is subtracted. This B feed-down contribution is evaluated using the B-meson
cross-section from FONLL \cite{bib:FONLL}, the ${\rm B} \rightarrow {\rm D+X}$ decay kinematics from EvtGen \cite{bib:evgen}, and the efficiencies for prompt and feed-down D mesons from simulations. FONLL calculations are used because they describe well the beauty hadron cross sections measured at the Tevatron \cite{bib:BprodTevatron} and at the LHC \cite{bib:BprodLHC_1}, \cite{bib:BprodLHC_2}. To account for possible
initial-state effects on the beauty-quark production, the resulting yield of D mesons from B feed-down is multiplied by a nuclear modification factor
$R_{\rm pPb}^{\rm feed-down}$ of D mesons from B decays. The central value for the calculation is chosen as $R_{\rm pPb}^{\rm feed-down}~=~R_{\rm pPb}^{\rm prompt}$ while the systematic uncertainty is evaluated by varying the hypothesis on the $\RpPb$ of feed-down D mesons in the range $0.9 < R_{\rm pPb}^{\rm feed-down}$/$R_{\rm pPb}^{\rm prompt} < 1.3$. 
This hypothesis is tuned on the basis of calculations obtained combining NLO pQCD cross sections from MNR \cite{bib:MNR}, EPS09 \cite{bib:EPS09} parameterizations of the nuclear parton distribution functions and B-decay kinematics from the EvtGen package \cite{bib:evgen}. The large uncertainty on the predictions at low $p_{\rm T}$ is taken into account.

The D-meson production cross section in pp collisions at $\sqrtsNN = 5.02$ TeV needed for the $\RpPb$ calculation is obtained by scaling the measured D meson production cross section at $\sqrt{s} = 7$ TeV using the ratio of FONLL predictions at $\sqrt{s} = 5.02$ and 7 TeV with the procedure described in \cite{bib:ppscaling}. The uncertainty on the scaling decreases with increasing $p_{\rm T}$ from 17\% to 3\% as estimated by varying the parameters entering the FONLL calculations, namely the factorization and renormalization scales and the charm quark mass. 
\begin{figure}[t]
\begin{center}
\includegraphics[width=14pc]{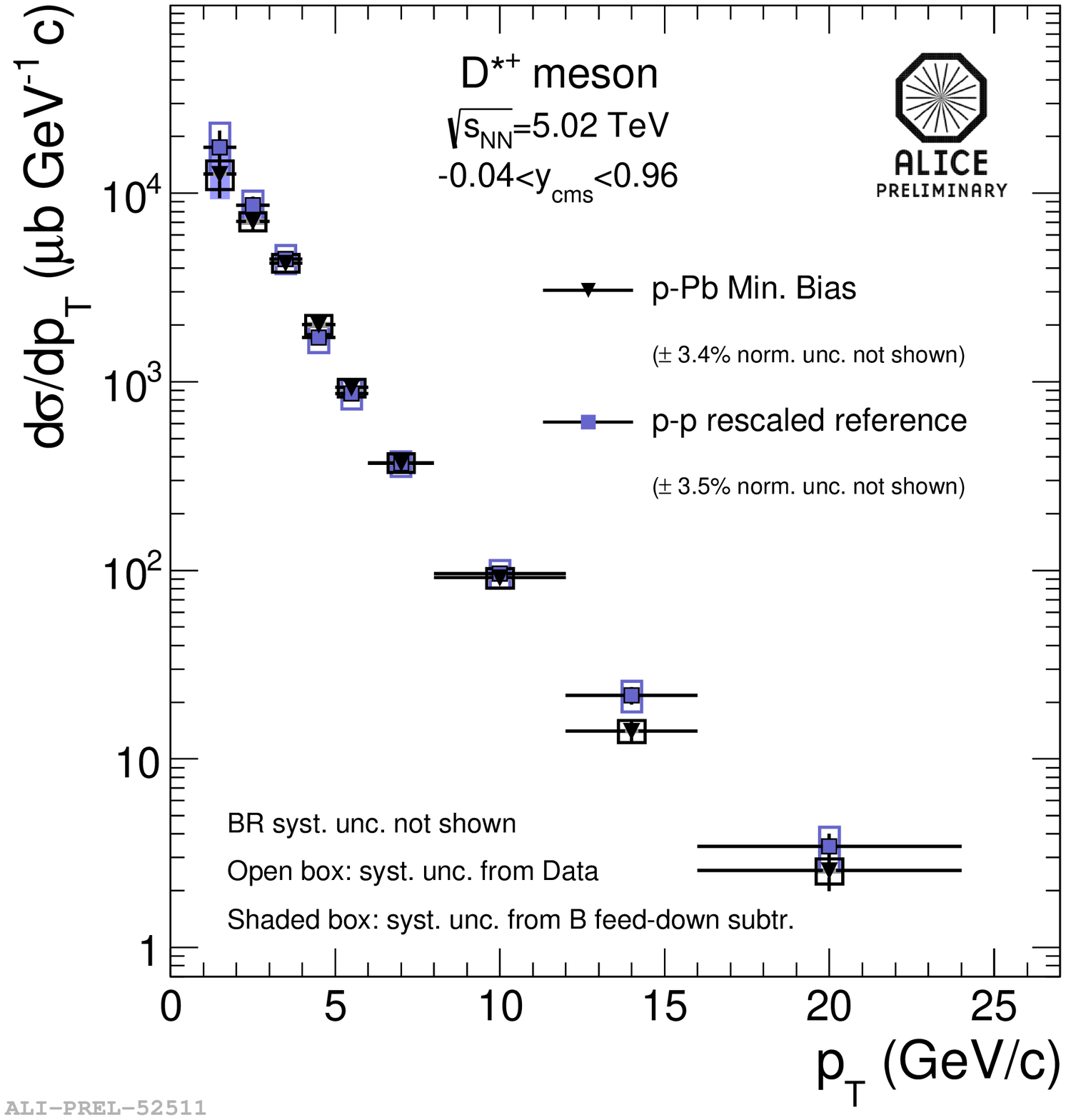}
\includegraphics[width=14pc]{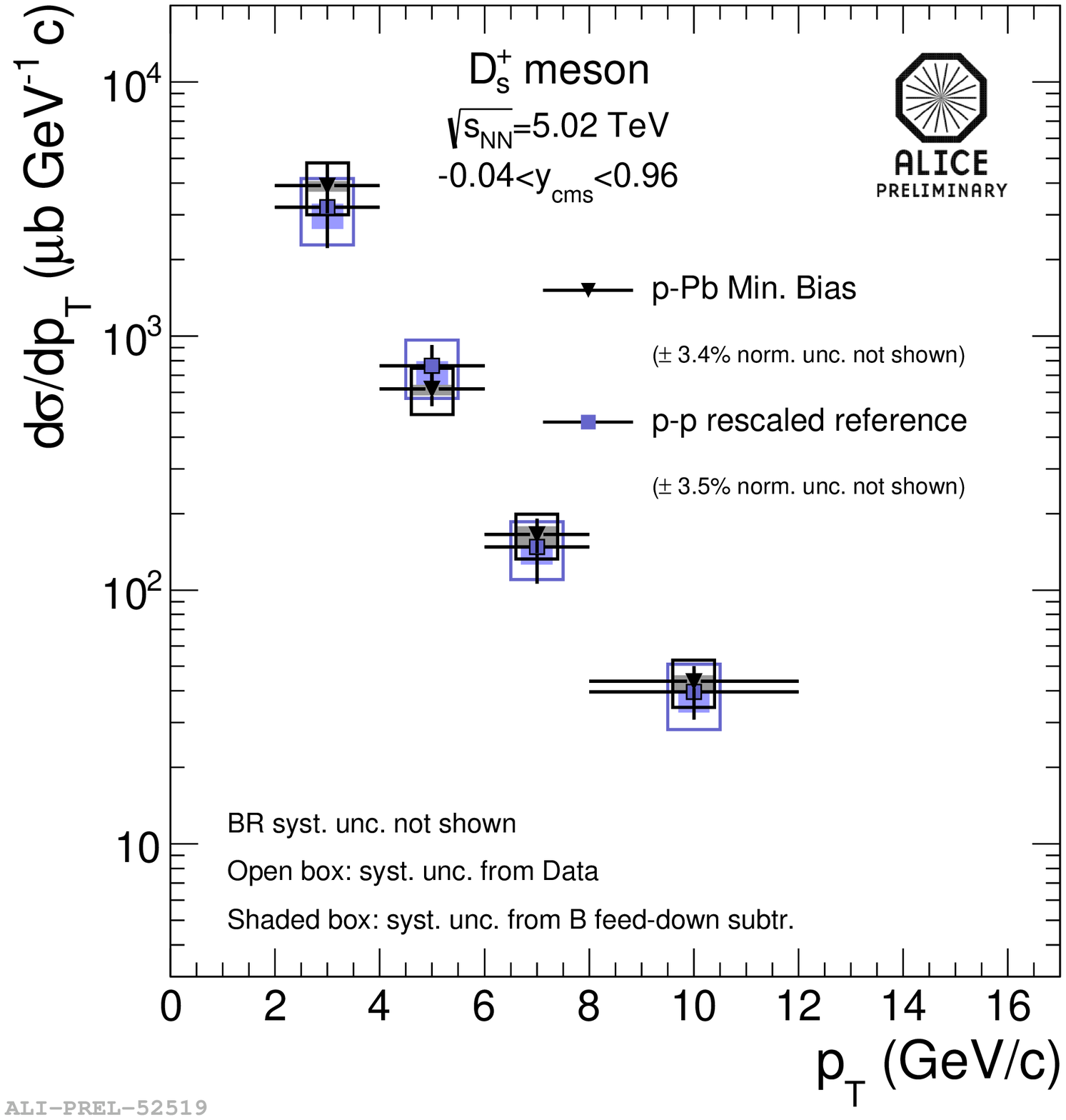}
\end{center}
\caption{\label{fig:Dstar_Dsigmadpt} $\Dstar$ (left) and $\Ds$ (right) $p_{\rm T}$-differential cross section for $\pPb$ collisions at $\sqrtsNN = 5.02$~TeV compared to the pp scaled cross sections. The shaded area represents the total systematic uncertainty.The 3.4\% normalization uncertainty is not shown.}
\end{figure}

Figure \ref{fig:Dstar_Dsigmadpt} shows the $\Dstar$ and the $\Ds$ cross sections in the momentum ranges $1< p_{\rm T} < 24$ GeV/$c$ and $2< p_{\rm T} < 12$ GeV/$c$ respectively, compared with the pp reference scaled by the number of nucleons in the lead nucleus $A$.
Several sources of uncertainties are considered. The systematic uncertainty on the yield extraction is determined in each $p_{\rm T}$ interval by repeating the fit in a different mass range, varying the background fit function (a parabola instead of an exponential for $\Dzero$, $\Dplus$, $\Ds$ and a power law multiplied by an exponential or a polynomial for the $\Dstar$) and by counting the histogram entries in the invariant mass region of the signal after subtracting the background estimated from the side bands. The effect of the imperfect implementation of the detector description in the Monte Carlo simulations is estimated by repeating the analysis with different selections on the track quality and on the decay topology. The uncertainty associated to the particle identification is studied by comparing the cross sections obtained with and without PID. The effect on the selection efficiency due to the shape of the simulated D meson spectrum is estimated from the relative difference between the Monte Carlo efficiencies obtained using different $p_{\rm T}$ shapes (Pythia, FONLL, flat  $p_{\rm T}$ distribution). The estimated systematic uncertainties vary from $\sim$15\% to $\sim$6\% depending on the $p_{\rm T}$ interval and the D-meson species. The systematic uncertainty on the feed-down subtraction was discussed above.

\section{D-meson nuclear modification factor}
\begin{figure}[!h]
\begin{center}
\includegraphics[width=0.36\textwidth]{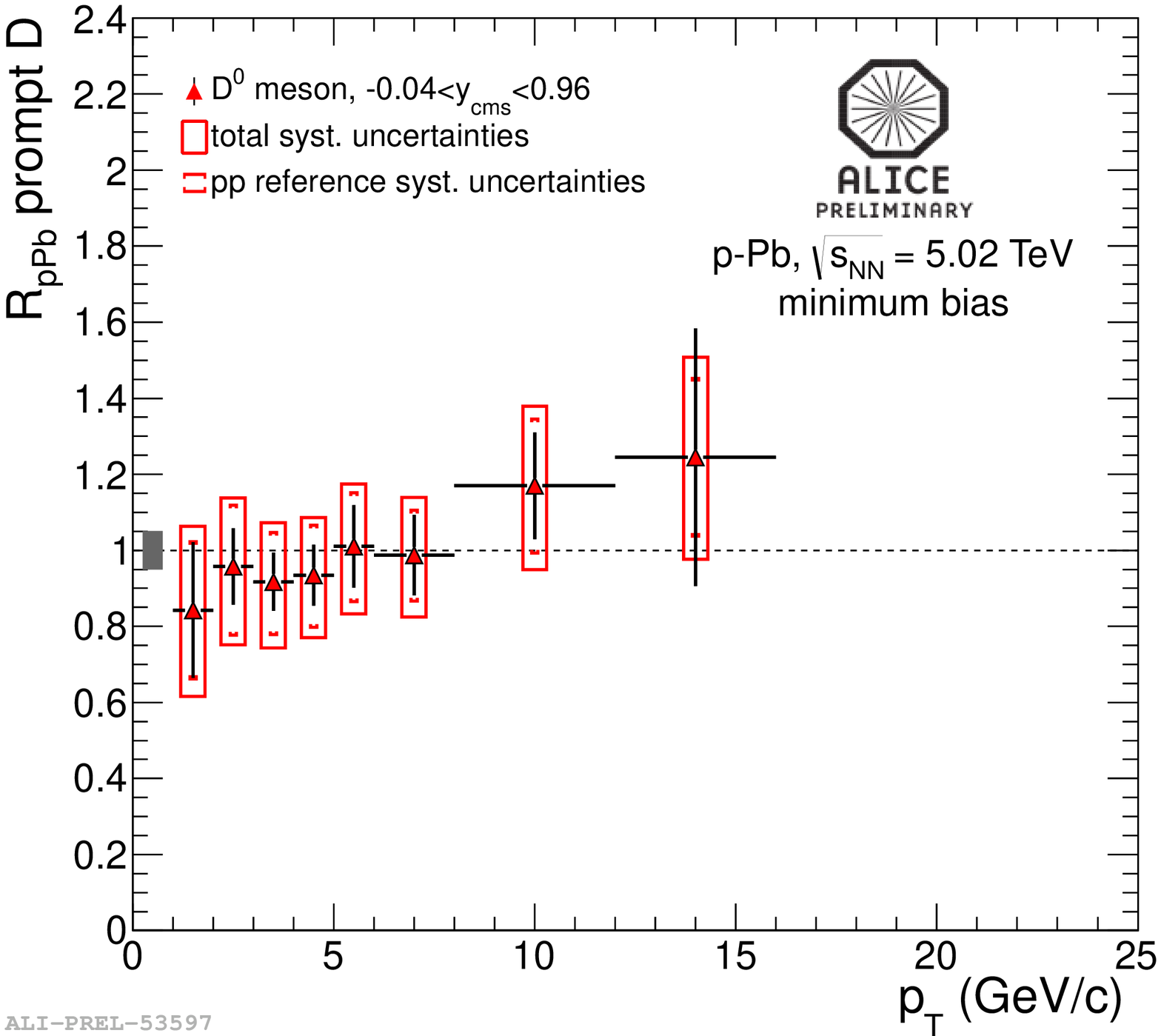}\hspace{1pc}
\includegraphics[width=0.36\textwidth]{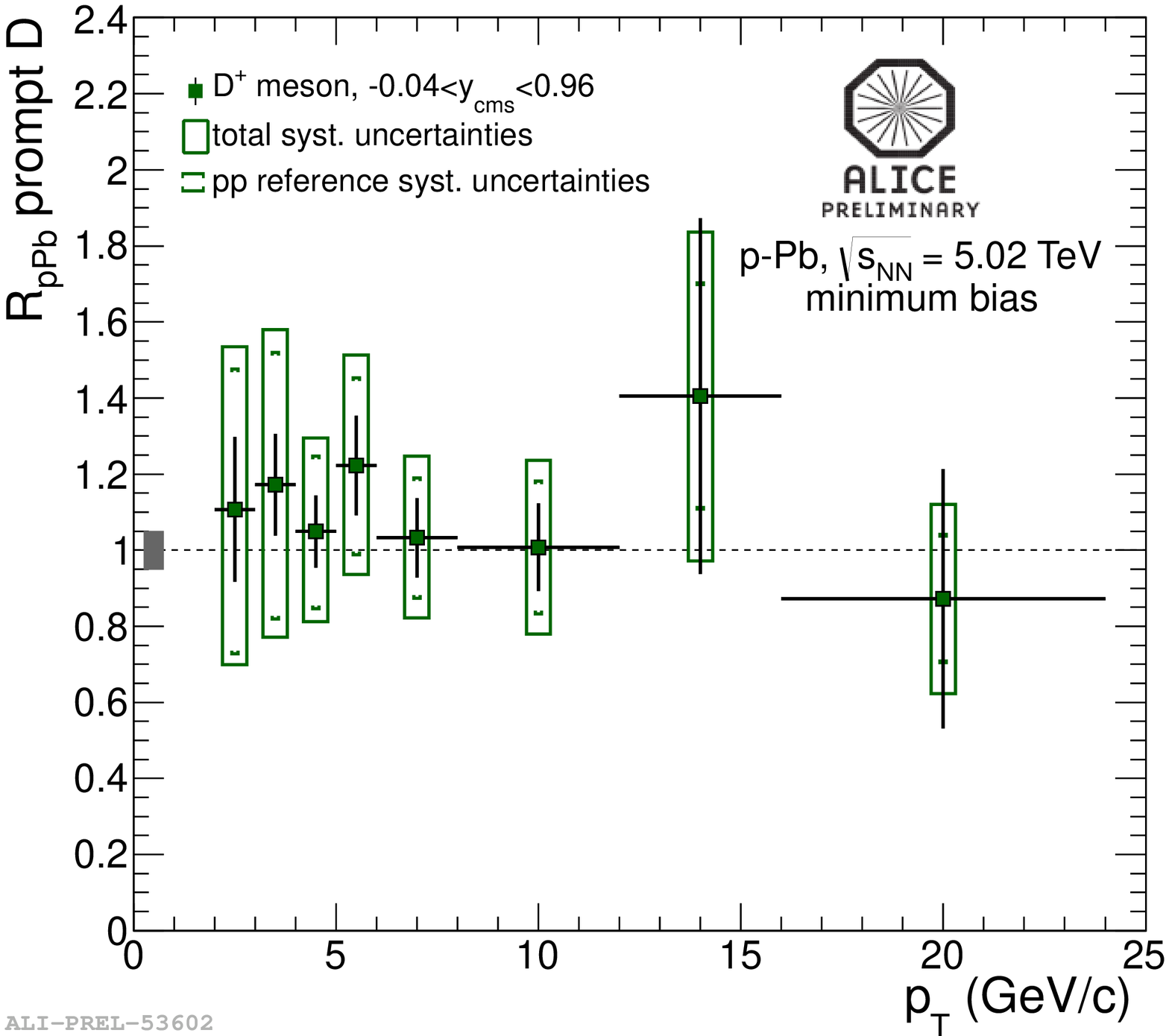}\\
\includegraphics[width=0.36\textwidth]{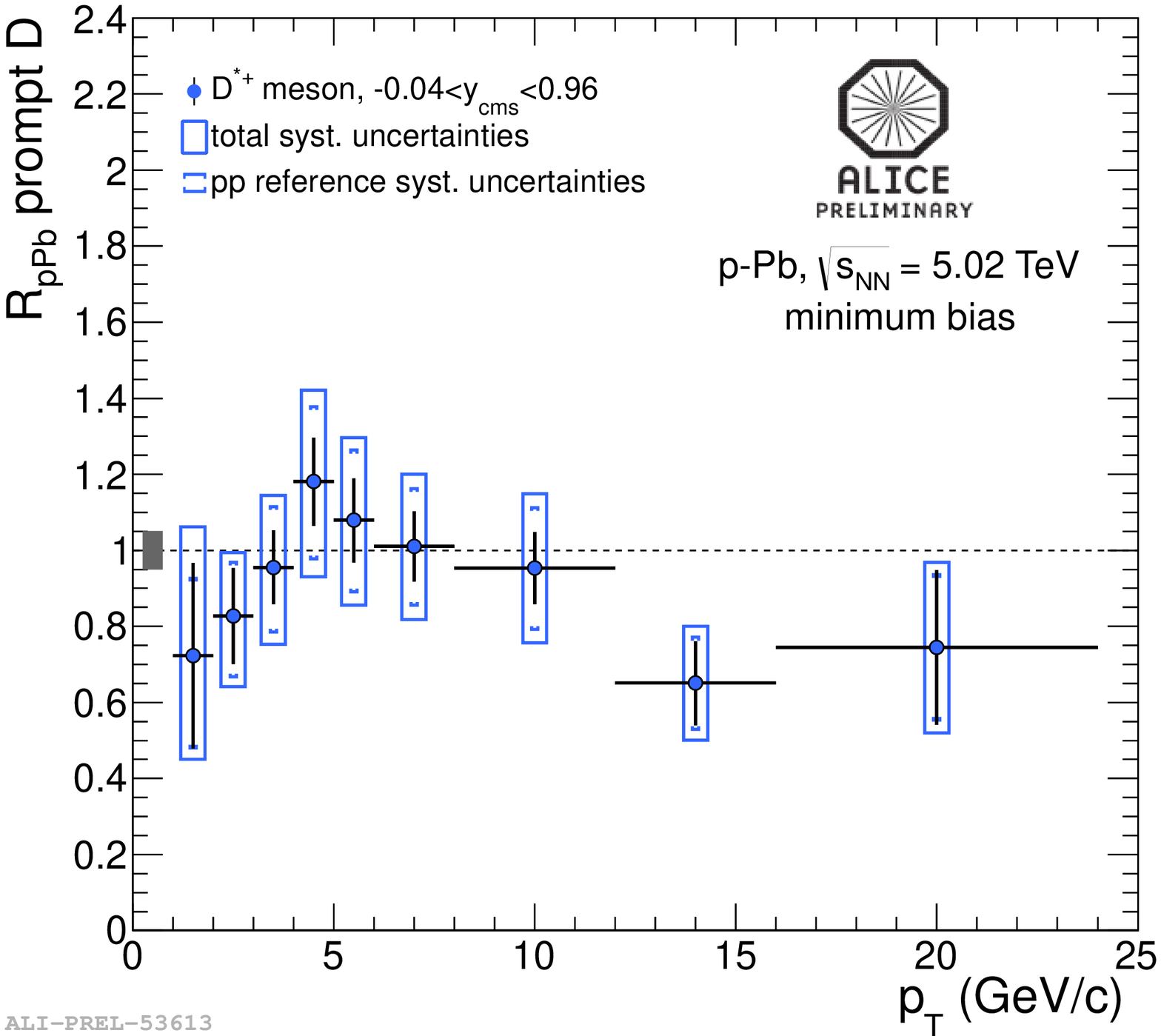}\hspace{1pc}
\includegraphics[width=0.36\textwidth]{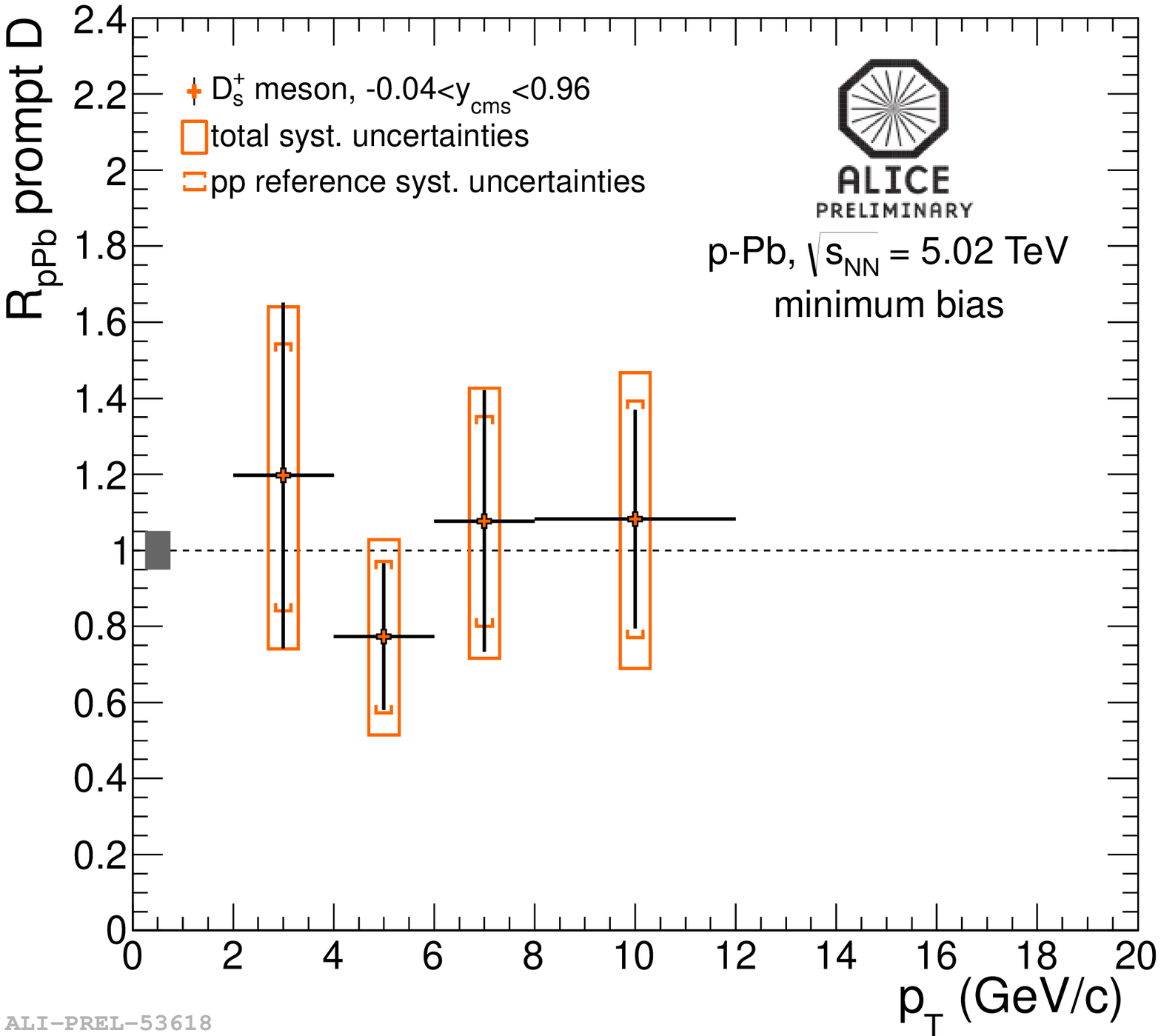}\\
\includegraphics[width=0.36\textwidth]{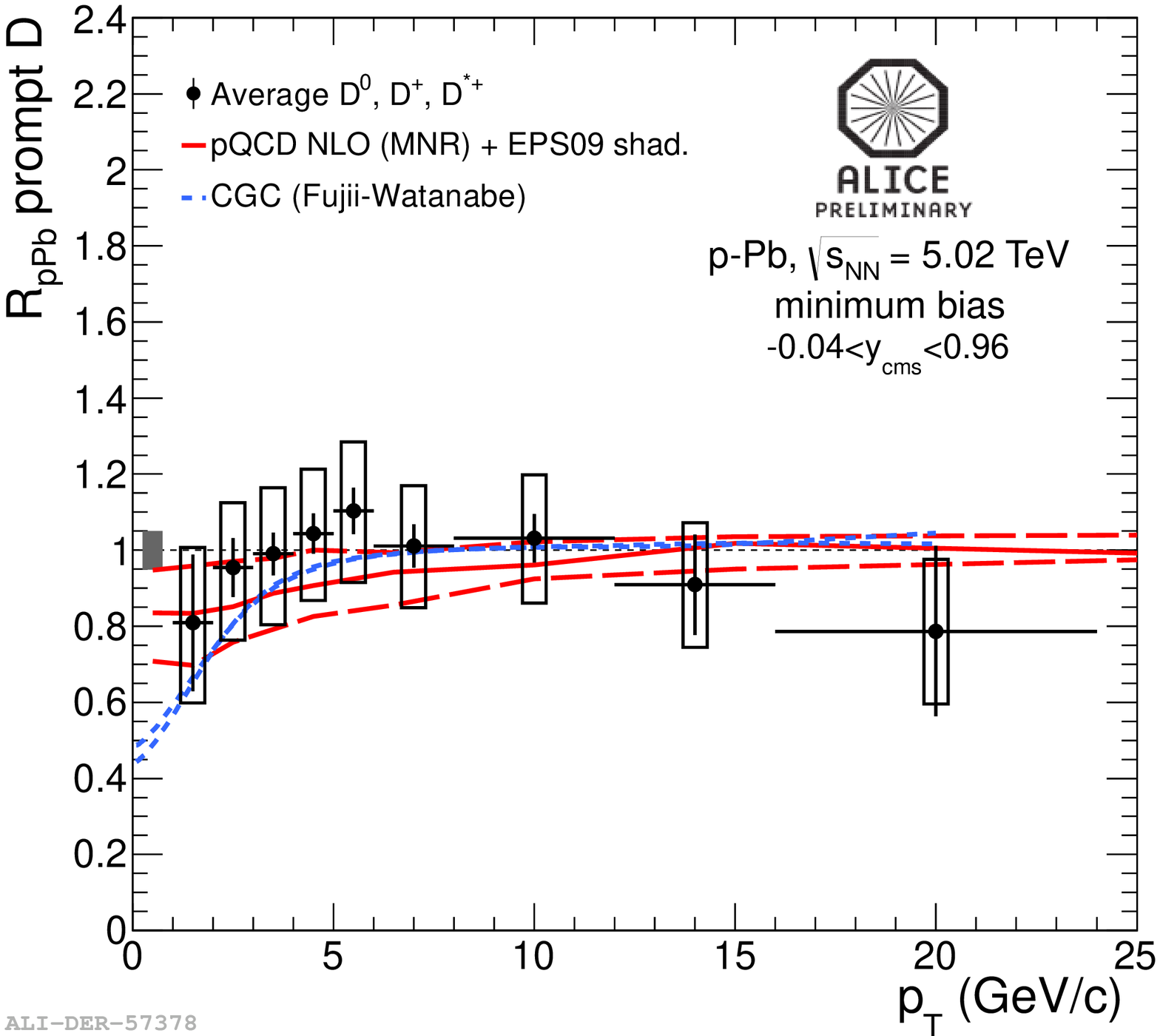}\hspace{1pc}%
\begin{minipage}[b]{14pc}\caption{\label{fig:RpPb_DzeroDplusDstarDs} $\Dzero$, $\Dplus$ (top plots), $\Dstar$, $\Ds$ (middle plots) $p_{\rm T}$-dependent $\RpPb$ at $\sqrtsNN~=~5.02$~TeV.\\ Bottom plot: average $\Dzero$, $\Dplus$, $\Dstar$ $\RpPb$ compared with theoretical models.}
\end{minipage}
\end{center}
\end{figure}
The $\pPb$ spectra are quantitatively compared to the pp reference by computing the nuclear modification factor, $\RpPb$. 
Figure \ref{fig:RpPb_DzeroDplusDstarDs} shows the $\RpPb$ of the $\Dzero$, $\Dplus$, $\Dstar$ and $\Ds$.
 The nuclear modification factors of the four meson species are in agreement with each other, and they are compatible with unity within the uncertainties in the full $p_{\rm T}$-range considered. 
This measurement suggests that the large suppression of the D-meson yield observed in $\PbPb$ collisions is mainly due to final-state effects.
The weighted average of the $\RpPb$ of $\Dzero$, $\Dplus$, $\Dstar$ in the $p_{\rm T}$ range $1<p_{\rm T} < 24$ GeV/$c$ is calculated using the statistical uncertainties as a weight. The systematic error on the average is calculated by propagating the uncertainties through the weighted average, where the contributions from tracking efficiency, B feed-down correction and scaling of the pp reference are taken as fully correlated among the three species. The bottom plot of figure \ref{fig:RpPb_DzeroDplusDstarDs} shows $\RpPb$ compared with pQCD calculations based on the MNR \cite{bib:MNR} code combined with nuclear parton distributions functions from EPS09 \cite{bib:EPS09} parameterizations and with a model based on the Color Glass Condensate \cite{bib:CGC2}. Both models describe the data within uncertainties.

\section{Conclusions}
The measurement of the  nuclear modification factor of prompt $\Dzero$, $\Dplus$, $\Dstar$ and $\Ds$ in $\pPb$ collisions at $\sqrtsNN = 5.02$ TeV is presented. The measured $\RpPb$ is compatible with unity within systematic uncertainties over the full $p_{\rm T}$ range exploited confirming that the strong suppression observed in central $\PbPb$ interactions can be interpreted as a final state effect due to in-medium parton energy loss. Theoretical models including nuclear shadowing or saturation effects can describe the data within uncertainties.

\end{document}